\documentstyle[aps,prd,twocolumn,epsf]{revtex} 
\newcommand{\be}{\begin{equation}} 
\newcommand{\ee}{\end{equation}} 
\newcommand{\beq}{\begin{eqnarray}} 
\newcommand{\eeq}{\end{eqnarray}} 

\begin{document} 
\twocolumn[
\begin{center}
{\large\bf\ignorespaces
The Gluon Propagator
without lattice Gribov copies on a finer lattice}\\
\bigskip
C.~Alexandrou \\
   {\small\it Department of Physics, University of Cyprus, P.O. Box 20537,
CY-1678 Nicosia, Cyprus} \\ 
\bigskip
Ph.~de~Forcrand \\
{\small\it Institut f\"ur Theoretische Physik, ETH H\"onggerberg,
 CH-8093 Z\"urich, Switzerland, and \\
CERN, Theory Division, CH-1211 Geneva 23, Switzerland}\\
\bigskip
E.~Follana \\
 {\small\it Department of Physics, University of Cyprus, P.O. Box 20537,
CY-1678 Nicosia, Cyprus} \\ 
\medskip
{\small\rm (Nov. 6, 2001)} \\
\bigskip
\begin{minipage}{5.5 true in} \small\quad
We extend our study of the gluon propagator in quenched  lattice QCD
 using the Laplacian gauge to a finer lattice. 
We verify the existence of a pole mass as we take the continuum limit
and deduce a value of $\sim 600^{+150}_{-30}$~MeV for this pole mass.
We find  a finite value of $(454(5)\>{\rm MeV})^{-2}$ 
for the renormalized zero-momentum
propagator, in agreement with results on coarser lattices.

\medskip

\end{minipage}
\end{center}
\vspace{6mm}
]

\section{Introduction}
The gluon propagator, although not an observable quantity, plays 
an important role
in phenomenological non perturbative studies. A framework for such studies 
 is provided, for instance, by  the Dyson Schwinger
equations (DSE) which recently have been applied, among other topics, 
to the study of the quark 
gluon-plasma~\cite{Roberts}. This application
is particularly important because  it complements
experimental activity at RHIC and also 
because of the difficulties of applying lattice QCD 
to the case of non-zero chemical potential, although 
some progress is being achieved in this direction~\cite{FK}.
However in the Dyson-Schwinger approach the study of the gluon propagator 
is still inconclusive because of the various truncations needed
to solve the coupled set of equations~\cite{Roberts}.
For instance, in the  ghost-free axial
gauge, many studies which
used a simplified version of the three-gluon vertex supported 
an infrared enhanced
gluon propagator of the form $1/q^4$.
Such a behaviour, driven by the  vacuum polarisation diagram,
  was disputed by other DSE studies~\cite{West}
as being due to a flaw in setting to zero the
 second scalar function that enters in the definition
of the gluon propagator.
Similar disputes also occur in the case of the Landau gauge where
some studies,  which assume dominance of the gluon-vacuum polarisation,
find infrared enhancement~\cite{Pennington} whereas
others~\cite{Stingl}, which use a rational polynomial Ansatz for the 
self energies and vertices, find an infrared vanishing 
propagator. After including the ghost propagator, recent studies 
in the Landau gauge favour an infrared-vanishing gluon propagator~\cite{AS}.

Lattice QCD provides a well defined theoretical framework for non perturbative
physics and it is well suited for the study of the gluon propagator.
  A series of papers \cite{Leinweber}, appeared over the past couple 
of years, provide a detailed study of the behaviour of the gluon propagator
in quenched lattice QCD in the Landau gauge. However, fixing to 
Landau gauge on the lattice is an iterative procedure 
which stops upon reaching any of a large number of local minima,
called ``Gribov copies''.
Since the global minimum itself cannot be reached, the effect of trading it
for a local minimum is largely unknown.
This uniqueness of the gauge condition
puts us in a good position
to obtain physical results on quantities like the pole mass which
one may expect to be gauge invariant,
and compare to the corresponding values used in
phenomenology. 
The issue of gauge invariance of the gluon pole mass
 was addressed in ref.~\cite{KKR}.
For a certain class of covariant gauges, the Ward identities that determine
the gauge dependence of QCD dispersion relations were derived. Using these
relations one can examine the gauge independence of the poles of the
gluon propagator. It was argued that if the loop expansion
holds, then a non-zero pole of the transverse gluon propagator is gauge
invariant.
 
In ref.~\cite{paperI}, we showed that
a good description of the gluon propagator was provided by an Ansatz
which admits a dynamically generated gluon mass~\cite{Cornwall} and thus 
points to an infrared regularised gluon propagator. By
analytic continuation to negative values of $q^2$ we obtained an estimate
of the pole mass. 
The existence of a gluon mass has important phenomenological
implications~\cite{Parisi}.
 Total cross sections in hadron-hadron collisions, proton-proton
elastic scattering  and diffractive phenomena can be well understood
if there is a finite correlation length
 for the gluon field~\cite{pomeron}. For instance
in the Pomeron exchange model of Landshoff and Nachtmann~\cite{LN} a gluon
propagator which is infrared finite is shown to eliminate
 the troublesome singularity
 in the Pomeron calculation of hadron-hadron scattering.
Whereas a  bare gluon mass would lead to problems with unitarity, a dynamically
generated mass  vanishing in the ultraviolet reproduces the correct
perturbative result for the gluon propagator and is consistent with unitarity.

It is the purpose of the present work to check the robustness of our earlier
results on the gluon propagator as we take the continuum limit.
We thus
extend our previous 
calculation on coarser lattices \cite{paperI} to a finer lattice
at $\beta=6.2$. In ref.~\cite{paperI} we  included 
a comparison of  results in different physical volumes demonstrating that,
for the quantities of interest here, such as  the pole mass and the zero
momentum limit of the gluon propagator, a lattice size of about 1.5~fm 
suffices. 
Therefore for this study we use a lattice of spatial size $\sim 1.7$~fm.
Our observable is the transverse part, $D(q^2)$, of the propagator.
The excellent scaling behaviour which we observe for  $D(q^2)$ 
enables us to extract accurately the change with $\beta$ of the lattice
spacing. 
In the infrared, it allows us to control the cutoff effects on the pole
mass and to study its continuum limit.

Our notation is the same as that of ref.~\cite{paperI} and 
we refer the reader to~\cite{paperI} for the details of our approach.

\section{Scaling}
Reasonable scaling was already observed 
 in \cite{paperI} where
we compared data at $\beta=5.8$ and $6.0$ on a lattice of size $16^3\times32$.
Here we compare $\beta=6.0$ and $\beta=6.2$.

\begin{figure}[bh]
\begin{center}
%\epsfxsize=5truecm
%\epsfysize=5truecm
%\mbox{\epsfbox{scaling.ps}}
\leavevmode
\epsfxsize=\columnwidth\epsfbox{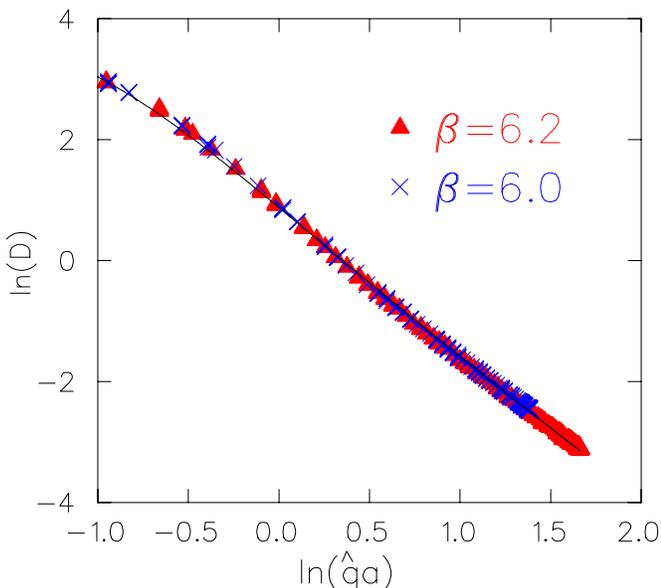}
%\vspace*{0.3cm}
\caption{Data at $\beta=6.0$ ($16^3\times32$ lattice; crosses)
 and at $\beta=6.2$
($24^3\times 48$ lattice; triangles)
 fall on a universal curve.}
\end{center}
\label{fig:scaling}
\end{figure}

The results at $\beta=6.2$ were obtained from    
 220 configurations generated  by the UKQCD collaboration  
on a lattice of size  $24^3\times 48$. At $\beta=6.0$ we used 200 
configurations of size $16^3\times 32$ from the NERSC archive~\cite{connection}.
Our Laplacian gauge condition consists of aligning along a fixed orientation
the local 3-color frame built from the two lowest-lying eigenvectors of the covariant Laplacian.
For implementation details, see \cite{paperI}.
%We measure the transverse gluon propagator $D(q^2)$.
Being now closer
to the continuum limit, we find very good scaling behaviour  for $D(q^2)$
as demonstrated
in Fig.~1, where the two data sets fall on the
same curve after applying the linear transformation
$$
{\rm ln}(D_{\beta=6.0}({\rm ln}qa_{\beta=6.0})) =
{\rm ln}(D_{\beta=6.2}({\rm ln}(qa_{\beta=6.2}-b))+c .
$$
The two fitted parameters take values
$b=0.277 \pm 0.022$ and $c=-0.574\pm 0.053$, which
yields the scaling ratios
\be
\frac{Z_{\beta=6.2}}{Z_{\beta=6.0}} = 1.02 \pm 0.14,  \hspace*{0.5cm}
\frac{a_{\beta=6.2}}{a_{\beta=6.0}} = 0.758 \pm 0.017.
\ee
This ratio of lattice spacings is consistent with that
 obtained from measurements
of the string tension~\cite{Bali}, and very close to the value $0.729$ obtained
from the interpolation formula of the Alpha-collaboration~\cite{Alpha} for $r_0/a$.
Therefore, we use for the lattice spacing the value
$a^{-1}(\beta=6.2)=2.718$~GeV of Ref.\cite{Bali},
corresponding to a string tension $\sqrt{\sigma}=440$~MeV.

The renormalised zero momentum propagator also exhibits good scaling.
We obtain a value of $(454(5) \rm{MeV})^{-2}$  in agreement
 with our previous value
of $\sim (445 \rm{MeV})^{-2}$, both for a   
  renormalization point of $1.943$ GeV.

\begin{figure}[h]
\begin{center}
\leavevmode
\epsfxsize=\columnwidth\epsfbox{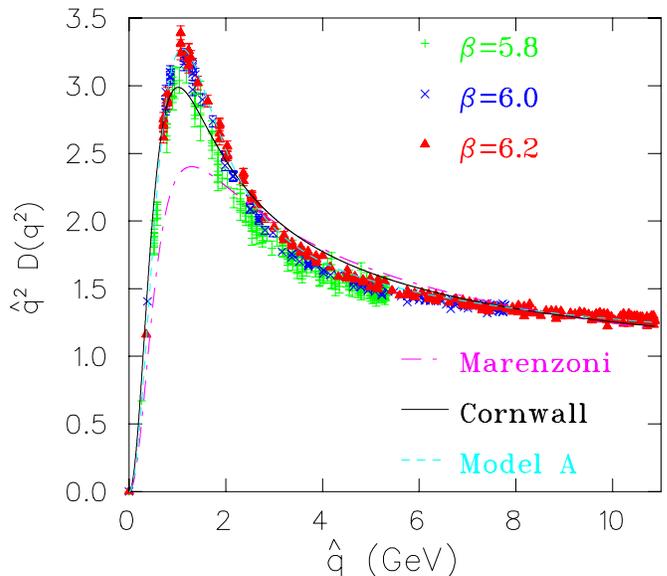}
\caption{The transverse gluon propagator multiplied by $\hat{q}^2$  
at $\beta=5.8, 6.0$ and  $6.2$ in physical units. 
The fits to three models are shown: 
Marenzoni (dashed dotted line)~[17],
%~\cite{Marenzoni},
Cornwall (solid line)~[10],
%~\cite{Cornwall}, 
and model A (dashed line) of ref.~[7]}
%\cite{Leinweber}.}
\label{fig:fits}
\end{center}
\end{figure}

Our new  transverse propagator data at $\beta=6.2$ are shown in 
Fig.~\ref{fig:fits} together with our previous results.
Since the $\beta=6.0$ and $6.2$ data agree so well over the whole 
momentum range, we can compare them
to a variety of continuum models.
We find that infrared enhancement like $(q^2)^{-2}$~\cite{Pennington},
as well as
Gribov-type infrared suppression of the form
$D(q^2) = Z q^2/(q^4+M^4)\> L(q^2,M^2) $~\cite{Gribov} or 
$ Z q^2/(q^4+2 a q^2+M^4) \> L(q^2,M^2)$~\cite{Stingl},  
are both excluded by our data.
Here, the function $L(q^2,M^2)$ enforces the
 perturbative ultraviolet behaviour of the propagator:
\be
L(q^2,M^2) = \biggl\{\frac{1}{2} \rm{ln}\left [
  (q^2+M^2)(q^{-2}+M^{-2})\right] \biggr\}^{-\frac{13}{22}} \quad.
\label{eqL}
\ee
The ansatz of Marenzoni et al.~\cite{Marenzoni}, 
\be
D(q^2) = Z \frac{M^{2\alpha}}{(q^2)^{1+\alpha} + (M^2)^{1+\alpha}} \quad,
\label{Marenzoni}
\ee
with a non-perturbative anomalous dimension $\alpha$,
gives a better description of the lattice data ($\chi^2/{\rm d.o.f}=18$)
but clearly underestimates the peak in Fig.~\ref{fig:fits}.
A much better description is provided by  model A of ref.~\cite{Leinweber}
\be
%\beq
D(q^2) = Z
\biggl[\frac{AM^{2\alpha}}{(q^2+M^2)^{1+\alpha}}
                + \frac{1}{q^2+M^2} \>L(q^2,M^2) \biggl]
% \nonumber \\
%L(q^2,M^2)&= &\biggl\{\frac{1}{2} \rm{ln}\left [
%  (q^2+M^2)(q^{-2}+M^{-2})\right] \biggr\}^{-13/22} \quad.
\label{modelA}
%\eeq
\ee
with $L(q^2,M^2)$ defined in (\ref{eqL}).
Although this model
 gives the  best fit to our data ($\chi^2/{\rm d.o.f}\sim 2$),
it requires one more parameter, and lacks a theoretical motivation.
Worse for us here, it cannot be
analytically continued.

Taking a different approach, 
Cornwall~\cite{Cornwall} allows for a dynamically generated gluon mass 
which vanishes at large momentum in accord with  perturbation theory:
$$
D(q^2)=Z\> \biggl[ \left(q^2+M^2(q^2)\right){\rm ln}
                             \frac{q^2+4M^2(q^2)}{\Lambda^2}\biggr]^{-1}
$$
with 
\be M(q^2)=M\left\{\frac{{\rm
    ln}\left[(q^2+4M^2)/\Lambda^2\right]}
                {{\rm ln}\left[4M^2/\Lambda^2\right]}\right\}^{-6/11} \quad.
\label{Cornwall}
\ee

As we already observed  on coarser lattices, we find that
also at $\beta=6.2$ Cornwall's Ansatz 
provides a reasonable fit to the data over the whole
momentum range ($\chi^2/{\rm d.o.f}=7$). We will thus keep it as one
possible way to extrapolate to negative values
of $q^2$ for the determination of the pole mass.

\section{Pole mass}
\begin{figure}[h]
\begin{center}
\leavevmode
\epsfxsize=\columnwidth\epsfbox{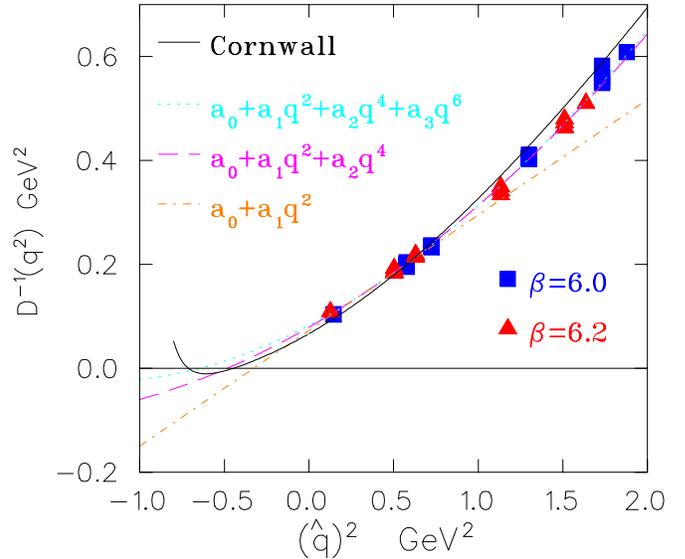}
\caption{Infrared inverse gluon propagator
at $\beta=6.0$ (squares) and $6.2$ (triangles).
The various extrapolations to negative $\hat{q}^2$ are: linear (dot-dashed), quadratic (dashed)
and cubic (dotted) polynomials,
and Cornwall's model (solid line). }
\label{fig:pole}
\end{center}
\end{figure}

As explained in the Introduction, a phenomenologically
 important question is whether the gluon propagator
has a pole mass. The pole is a zero of the inverse propagator $D^{-1}(q^2)$.
We show this quantity in physical units on 
Fig.~\ref{fig:pole}, combining our data at $\beta=6.2$ and $\beta=6.0$.
To determine the pole mass from $D^{-1}(q^2)=0$, an analytic
continuation to negative $q^2$ is needed. Given our finite amount of data,
this extrapolation procedure is ambiguous. It is essential to test 
the robustness of a possible pole by comparing a variety of plausible
extrapolations. Fortunately, the deviation of $D^{-1}(q^2)$ from a
linear function of $q^2$
in the infrared
is small
(much smaller than in Landau gauge~\cite{paperI}),
which increases confidence in the extrapolation.
We consider here linear, quadratic and cubic
polynomials in $q^2$, as well as Cornwall's ansatz (\ref{Cornwall}).
All four choices indicate the presence of a pole.
The variation in its location gives us crude systematic error estimates.

Given the curvature of the data, a linear fit provides a lower bound
for the pole mass, with a value of $505$~MeV. This mass increases to
$693(20)$~MeV with a quadratic fit, and reaches $800$~MeV for a cubic
fit. Note that the fitted coefficients of the higher powers of $q^2$ 
keep decreasing, which indicates the soundness of an extrapolation
based on a Taylor expansion. However our data are not sufficient to
reliably fit higher-degree polynomials: in particular, the results
depend on the fitted momentum interval and on the presence or absence
of data removed by the ``cylindrical cut'' of ref.~\cite{Leinweber},
whose purpose is to eliminate momenta most affected by lattice artifacts.
In comparison with these simple polynomial fits, Cornwall's ansatz gives 
a pole mass of $669(6)$~MeV, consistent with that of the quadratic polynomial.
The various extrapolations are shown together in Fig.~\ref{fig:pole}.

\begin{figure}
\begin{center}
\leavevmode
\epsfxsize=\columnwidth\epsfbox{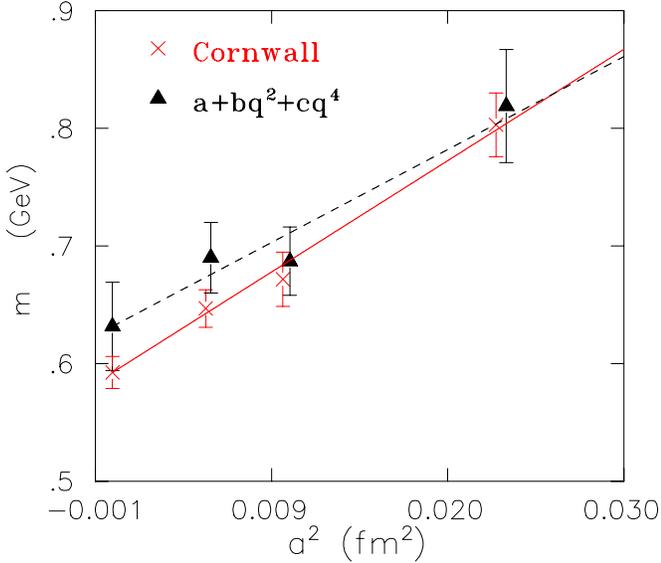}
\vspace*{0.2cm}
\caption{The pole mass at $\beta=5.8, 6.0$ and $6.2$
extracted from fitting to  a quadratic polynomial (data shifted to the right for
clarity)
 and to Cornwall's Ansatz, as a function of the lattice spacing $a^2$. 
The continuum values
are obtained by linear extrapolation in $a^2$. }
\label{fig:mpole}
\end{center}
\end{figure}

By performing the same analysis at the three values of the lattice spacing 
$a(\beta), \beta=5.8, 6.0$ and $6.2$
we have studied,
we can extrapolate the pole mass to the continuum limit.
In Fig.~\ref{fig:mpole}, we compare such an extrapolation in $a^2$, for pole masses obtained
by fitting a quadratic polynomial or Cornwall's ansatz.
It can be seen that both give consistent results, at fixed lattice spacing
as well as in the continuum.
The continuum values are $ 632(38)$~MeV and $592(14)$~MeV using
 the quadratic polynomial and Cornwall's Ansatz respectively.

The reasonable robustness of our analysis, with respect to a change
in the lattice spacing as well as in the analytic continuation procedure,
leads us to conclude that there is strong evidence that a pole
exists in the gluon propagator, with a mass of about $600$~MeV.
Because of the curvature of the
inverse propagator, the systematic error in the 
extrapolation to negative $q^2$ is asymmetric.
From the variation observed above with the choice of analytic continuation,
we estimate it at $\sim -30, +150$ MeV.
Based on our study of finite volume effects~\cite{paperI}, and
given our present lattice size, we expect
negligible finite size corrections to this pole mass.
Comparing it 
to the glueball mass of $1.73(0.05)(0.08)$~GeV\cite{Morningstar},
it appears close to one third of the glueball mass
(rather than one half as sometimes speculated).
A pole mass of $500-800$~MeV  is also within the range
needed to fit
experimental data in various phenomenological studies~\cite{Parisi,pomeron}.

\begin{figure}
\begin{center}
%\epsfxsize=10truecm
%\epsfysize=10truecm
%\mbox{\epsfbox{Dt.ps}}
\leavevmode
\epsfxsize=\columnwidth\epsfbox{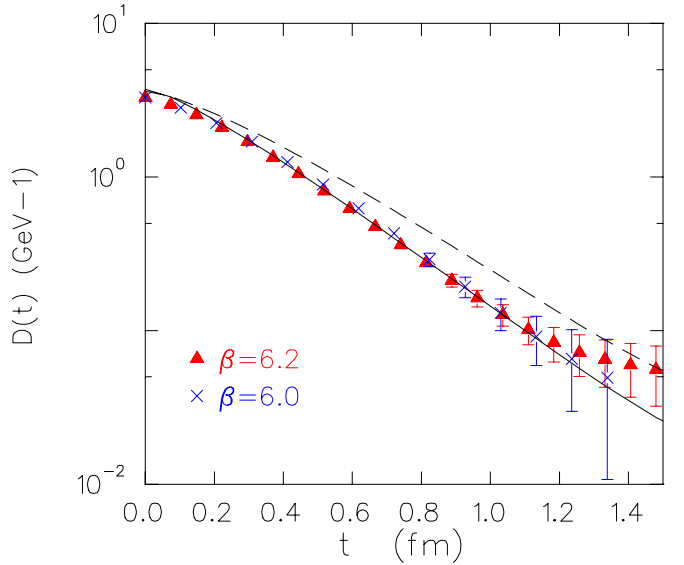}
\vspace*{0.2cm}
\caption{Time-slice gluon correlator at $\beta=6.0$
and $6.2$. 
The dashed line shows Cornwall's model fitted to $D^{-1}(\hat{q}^2)$ after
the cylindrical momentum cut; the solid line is a direct fit of the
same model to the
time-slice correlators, excluding the first few time-slices.}
\label{fig:Dt}
\end{center}
\end{figure}

As a further, model-independent check on the value of the pole mass,
we measured  
the correlator of the gluon field averaged over a time-slice, namely~\cite{paperI}
\be
C(t) =\frac{1}{L_s^3} \frac{1}{8} \sum_{a=1}^8 \frac{1}{3} \sum_{\mu=1}^3 
( \sum_x^{L_s^3} A_\mu^a(\vec{x},0) )~( \sum_x^{L_s^3} A_\mu^a(\vec{x},t) )
\label{Ct}
\ee
The exponential decay, if at all, of this correlator at large time
is governed by the pole mass. Therefore, one can in principle measure the
pole mass in a model-independent way.
However, statistical noise limits the usefulness of this approach.
We display the correlator in Fig.\ref{fig:Dt}.
The data is of insufficient quality to measure an exponential fall-off,
and a fitting procedure is needed.
Although this correlator is measured
on the same configurations as $D(q^2)$, the various momenta are
given a different weight, so that
a fit to $C(t)$ will give different results than a fit to $D^{-1}(q^2)$, especially
after the cylindrical momentum cut in the latter. 
Since all momenta enter in $C(t)$, we fit the only ansatz which describes the gluon
propagator reasonably well over the whole momentum range, namely Cornwall's model,
directly to $C(t)$ instead of $D^{-1}(q^2)$ as before. 
The dashed line in
Fig.\ref{fig:Dt} shows the original fit of Cornwall's ansatz 
to $D^{-1}(q^2)$, which  already provides a fair description of the data. 
The solid line represents a direct fit of
the same 3-parameter ansatz to $C(t)$, excluding the first few time-slices
which are contaminated by contributions from excited states.
A simultaneous fit of the time-slice correlator data at $\beta=6.2$ and $6.0$
yields a pole mass of $739(81)$~MeV, in agreement with the value $669(6)$~MeV extracted
from $D^{-1}(q^2)$, but with
much larger statistical errors.

If one attempts a model-independent determination of the pole
mass from the effective mass $m_{\rm{eff}}(t) = -Ln(C(t+1)/C(t))$,
one obtains a value of $702(163)$~MeV, poorly determined but consistent with the direct fit.
Therefore, our different analyses give pole masses ranging from
$592$~MeV (continuum extrapolation of pole of propagator fitted by
Cornwall's Ansatz) to $739$~MeV (fit of time-slice correlator).
Taking into account the asymmetry of potential systematic errors,
we estimate the gluon pole mass to be $\sim 600^{+150}_{-30}$~MeV.

\section{Conclusions}

We have extended a previous lattice study of the gluon propagator in the Laplacian
gauge to a finer lattice and found good scaling behaviour. 
We confirm the existence of a pole 
as we approach the continuum limit. Applying a variety of different fits
we extract a pole mass in the range of $600 -30 +150$~MeV
in accord with the value found in phenomenological studies
for the description of hadron-hadron scattering. 

It would be very interesting to substantiate the gauge-invariance of 
the gluon propagator pole mass by similar studies in other gauges.
Ref.~\cite{Phil} proposes a non-local, gauge-invariant gluon propagator
based on the (Coulomb-like) Laplacian gauge in three dimensions.
It is argued there that the pole mass of this propagator
determines  the vector- pseudoscalar- mass splitting 
$M_V-M_S$ in heavy quarkonia. This relation is consistent with the
$SU(2)$ numerical results presented. For QCD, using the experimental values
for the splitting in $c\bar{c}$ and $b\bar{b}$ systems,
 the implication is that the pole
mass is $\sim 420$~MeV.
This is somewhat low compared to our estimate.
In Landau gauge, our first attempt~\cite{paperI} showed no indication
for a gluon propagator pole. However, lattice Gribov copies might play
a crucial role there. Moreover, Landau gauge appears more sensitive
than Laplacian gauge to finite-size effects.
 Thus this question requires
larger lattices as studied in~\cite{Leinweber}.

\vspace*{0.3cm}

\noindent
\underline{Acknowledgements:}
The $16^3\times 32$ lattice configurations came from the
Gauge Connection archive~\cite{connection}, and
the $24^3\times 48$ were provided by the UKQCD collaboration.

\end{document}